\documentstyle[11pt,newpasp,twoside,epsf]{article}
\def\etal{et al.~}

\def\edcomment#1{\iffalse\marginpar{\raggedright\sl#1\/}\else\relax\fi}
\reversemarginpar

\begin{document}
\title{A New Technique to improve RFI suppression in Radio Interferometers}

\author{D. Anish Roshi}
\affil{National Radio Astronomy Observatory (NRAO), Box 2, Green Bank,
West Virginia 24944, USA}

\author{R. A. Perley}
\affil{National Radio Astronomy Observatory (NRAO), P.O. Box O, 1003 Lopezville Road
Socorro, NM 87801-0387}  

\begin{abstract}
Radio interferometric observations are less susceptible to radio frequency 
interference (RFI) than single dish observations. This is primarily due to : (1) 
fringe-frequency averaging at the correlator output and (2) bandwidth decorrelation 
of broadband RFI. Here, we propose a new technique to improve RFI suppression of 
interferometers by replacing the fringe-frequency averaging process with a 
different filtering process. In the digital implementation of the correlator,
such a filter should have cutoff frequencies $< 10^{-6}$ times the frequency  at
which the baseband signals are sampled.  We show that filters with  
such cutoff frequencies and attenuation $>$ 40 dB at frequencies above
the cutoff frequency can be realized using multirate filtering techniques. 
Simulation of a two element interferometer system
with correlator using multirate filters shows that the RFI suppression at the output of 
the correlator can be improved by 40 dB or more compared to correlators using 
a simple averaging process. 
\end{abstract}

\section{Introduction}

Radio interferometers provide high-resolution, high-dynamic range radio images 
of celestial objects. 
It is well known that radio interferometers are less susceptible to radio frequency interference (RFI) 
compared to single dish. However, the dynamic range of the images in 
many cases is limited by interference indicating the need to improve 
RFI suppression. 

In this paper, we describe a new technique to improve RFI suppression in 
interferometers \footnote
{To our knowledge a technique similar to that discussed in this paper was used
earlier in Mauritius Radio Telescope data reduction (Sachdev, Udayashankar 1995).}
. The basic idea is to replace the averaging 
done at the output of the correlator with a multirate filtering process.  
A brief discussion of RFI suppression in existing interferometers is given
in Sec. 2. Sec. 3 describes the proposed
technique, gives the design of the multirate filter and presents simulation
results.  The limitations of the proposed technique are discussed in
Sec. 4 and the effects of the multirate filtering process in
the visibility and spatial domain are discussed in Sec. 5.

\section{RFI suppression in existing Radio Interferometers}
\label{secRFIim}

\subsection{Correlated and Uncorrelated RFI}

\begin{figure}
\plotone{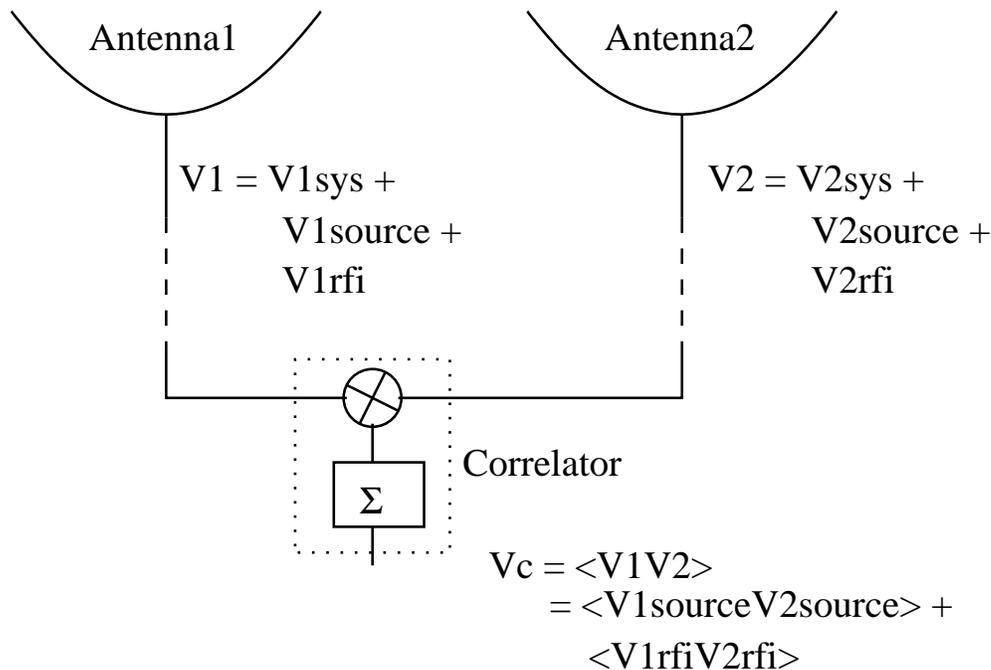}
\caption{Schematic of a two element interferometer near an RFI source. The 
outputs of the antennas consists of the RFI ($V_{1rfi}$, $V_{2rfi}$), 
the system noise ($V_{1sys}$, $V_{2sys}$) and the source 
noise ($V_{1source}$, $V_{2source}$). Interferometers usually use
a complex correlator to measure both amplitude and phase of the
output of the correlator. Here, for simplicity, we consider
the `real' part of the complex correlator. The mean value of the correlator 
output will be the sum of the time averages of the product of the
voltages due to the source and that due to the RFI. 
\label{fig1}
}
\end{figure}

In the case of interferometric observations interference can be broadly classified as 
correlated and uncorrelated RFI.
Consider a two element interferometer near a RFI source (Fig. 1). 
The voltages $V_1$ and  $V_2$ at the
outputs of the two antennas each consists of three components --- the system noise 
($V_{1sys}$, $V_{2sys}$), the source noise\footnote{The source noise is 
that component of the sky noise which is not `resolved'.} 
($V_{1source}$, $V_{2source}$) and the RFI component ($V_{1rfi}$, $V_{2rfi}$).
For the moment we will not consider fringe stopping, which is the case when the source is
at the north pole. Also we consider that the RFI is not moving with respect to the interferometer.
Interferometers usually use complex correlators, which are constituted of `real' and `imaginary'
parts. For the discussion in this paper, we consider a `real' correlator. But the analysis
is equally applicable to the `imaginary' part of the correlator as well.
The mean value of the correlator output, $V_c$, is given by
\begin{equation}
 V_c = <V_1 V_2> = <V_{1source}V_{2source}> + <V_{1rfi}V_{2rfi}>,
\end{equation}
where $<>$ denotes time average.
If the separation between antennas (baseline length) is $>$ a few 100 kms, which 
is the case for the VLBI (very long baseline interferometry), and if the RFI
source is near one of the antennas, then the second antenna will not 
pick up the RFI (ie say $V_{1rfi} = $ 0). This means that $V_c$ does not 
have any contribution from the RFI source or that the RFI is uncorrelated.
If this condition is not true then the RFI is correlated (for this
discussion we consider that the bandwidth of the system is very small
so that there is no bandwidth decorrelation). In most 
connected interferometers RFI is correlated in 
many baselines. For the VLBI, most terrestrial RFI is uncorrelated, but
satellite interference can be correlated in most `short' baselines. The 
technique discussed
in this paper to suppress interference deals with correlated RFI. Note
that the RMS (root mean square) value of the correlator output is 
proportional to the geometric mean of the
total powers at the inputs of the correlator if the source noise power
is much smaller than the system noise power.  
The correlator input powers have contributions from the RFI even if it is uncorrelated.
Thus uncorrelated RFI also limits the dynamic range
of the image. 
 
\subsection{Fringe stopping and correlated RFI}

\begin{figure}
\plotone{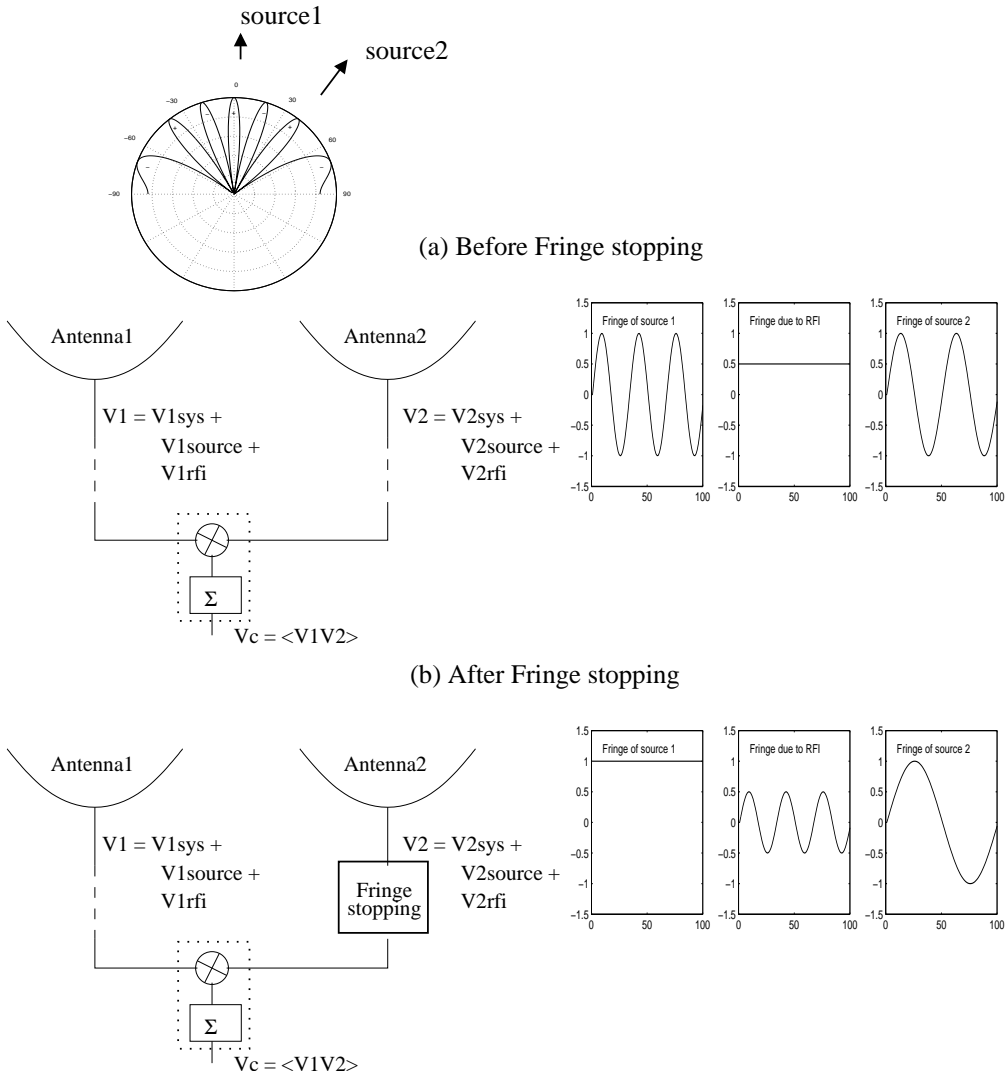}
\caption{Schematic showing the effect of fringe stopping in a two element
interferometer. The response of the two element interferometer is shown
on the top in polar diagram. If no fringe stopping is done the source
moves through this response producing a fringe at the output of the
correlator. The fringe frequency due to sources separated in sky 
are different which is schematically shown in figure (a). The RFI
power at the output of the correlator in this case will be constant.
After fringe stopping, as shown in figure (b), the component of the
correlator output due to the source at the phase center (source 1 in figure) 
will be constant and the RFI will produce a fringe. A source away
from the phase center (source 2) will still produce a fringe but its
frequency will be the difference 
between its original fringe frequency
and the fringe stopping frequency. The figure represents the case when
either the RFI is narrow band or the spectral resolution of the receiver
system is less than the differential fringe frequency.
\label{fig2}
}
\end{figure}

For simplicity consider that the baseline of the two element interferometer
is in the East-West direction. The output of 
the two antennas are connected to a correlator with cables of equal length 
(Fig. 2). The response of the interferometer $P(\theta)$ is 
given by (Thompson, Moran, Swenson, 2001)
\begin{equation}
P(\theta) = P_{ant}(\theta)\ \mbox{cos} \left(\frac{2\pi D \ \mbox{sin}(\theta)}{\lambda} \right),
\end{equation}
where $P_{ant}(\theta)$ is the cross response of the two antennas, $D$
is the baseline length and $\lambda$ is the operating wavelength. $\theta$ 
is measured with respect to the zenith, which we equate to 
the hour angle ($HA$) in radians for simplicity. A source moving through the interferometer response 
produces ``fringes'' at the output of the correlator. The fringe frequency
$f_{fringe}$ in the present case is given by,
\begin{equation}
f_{fringe} = \frac{D}{\lambda} \mbox{cos}(HA) \frac{\mbox{d}(HA)}{\mbox{d}t} \mbox{cos}(\delta),
\end{equation}
where $\delta$ is the declination of the source and $\frac{\mbox{d}(HA)}{\mbox{d}t}$ is
in units of radians per sec.
The general equation for the fringe frequency is (Thompson \etal 2001)
\begin{equation}
f_{fringe} = \frac{\mbox{d}(HA)}{\mbox{d}t} \ u \ \mbox{cos}(\delta),
\end{equation}
where $u$ is the projected baseline in the East-West direction in units of $\lambda$.
The above equation shows that the fringe frequency due to a source depends
on the projected baseline and earth's rotation 
rate ($= \frac{\mbox{d}(HA)}{\mbox{d}t}$). Now consider a terrestrial, stationary, correlated RFI.
Since the position of the RFI does not change with respect to the interferometer,
there will be no fringe due to the interference.

For observations with the interferometer, the interferometer is `phased' in the direction
of the interested radio source through out the observations. This process of
phasing the interferometer (which includes both delay compensation and fractional
phase compensation) is called `fringe stopping'. Fringe stopping essentially
makes the correlator output constant due to a (point) source at the phase center.
Sources away from the phase center produce fringes with frequencies equal to the
difference between $f_{fringe}$  of the source at the phase center and    
that of the source away from the phase center. Thus, in the above case, 
larger the difference between the $HA$ of a source and that at the phase
center the larger will be its `differential' fringe frequency.
This also means that the correlated RFI now produces a fringe with 
a frequency equal to $f_{fringe}$ of the source at the phase center because of 
fringe stopping.

\subsection{Fringe averaging and RFI suppression}

Usually in interferometric observations, one is interested in mapping sources within the
primary beam of the antenna. 
Typically the differential fringe frequencies due to these sources
is much smaller than $f_{fringe}$ of the source at the phase center.
The maximum differential fringe rate is given by
\begin{equation}
\Delta f_{fringe} \sim \frac{\mbox{d}(HA)}{\mbox{d}t} \frac{\Omega_{antenna}}{\Omega_{synth}} \sim  
                        \frac{\mbox{d}(HA)}{\mbox{d}t} \frac{D}{D_{antenna}},
\end{equation} 
where $\Omega_{antenna}$ and $D_{antenna}$ are the beam and diameter of each element of an interferometer,
$\Omega_{synth} \sim \lambda/D$ is the synthesized beam, $\lambda$ is the operating wavelength and $D$
is the baseline length. Consider, for example, 
$D$ = 10 km and $D_{antenna}$ = 25 m. The differential fringe frequency is $\sim$ 0.005 Hz
while $f_{fringe}$ of the source at the phase center is $\sim$ 3.5 Hz, for $\lambda = $ 0.21 m.
The low differential fringe frequency allows one to average the output
of the correlator for several seconds. Since the RFI produces a fringe
(due to fringe stopping process) this averaging reduces the amplitude of the
RFI at the correlator output thus making the interferometer less susceptible to
RFI (see Perley 2002 for a quantitative estimate of RFI attenuation in 
Very Large Array due to fringe averaging). An additional 
suppression of broadband RFI occurs
due to bandwidth decorrelation, which is a bonus over the former effect.
We will not discuss the bandwidth decorrelation effect any further here (see Thompson \etal 2001). 

\section{A new technique to improve RFI suppression in Interferometers}
\label{sectech}

The averaging at the correlator output can be thought of as a moving average filtering
process. The response of such a filter is shown is Fig. 4. The stop-band 
attenuation (i.e. the attenuation of frequencies beyond the cutoff frequency)
of the filter determines the amount of RFI suppression achieved in
this process if the fringe due to RFI is beyond the cutoff frequency of the 
filter. As seen from Fig. 4, stop-band attenuation in the worst case
is 13.5 dB. We propose that RFI suppression in interferometers can be improved by
using a filter with high stop-band attenuation at the output of the correlator.
In the following sections we show how such filters can be designed and present  
results of the simulation done to tests its performance.

\subsection{Multirate filtering}

\begin{figure}
\plottwo{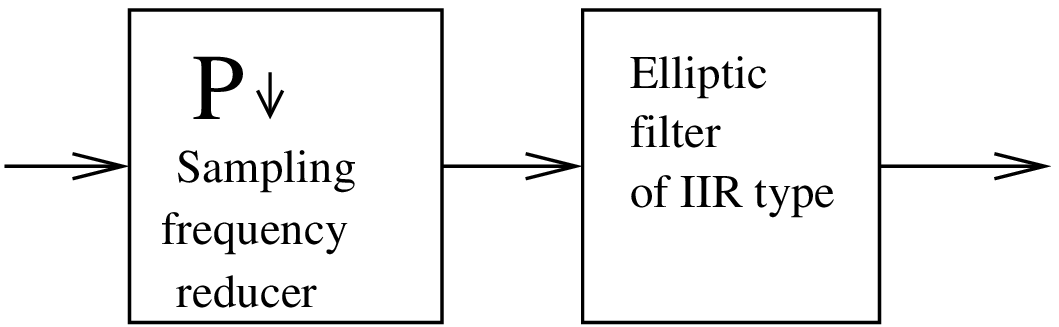}{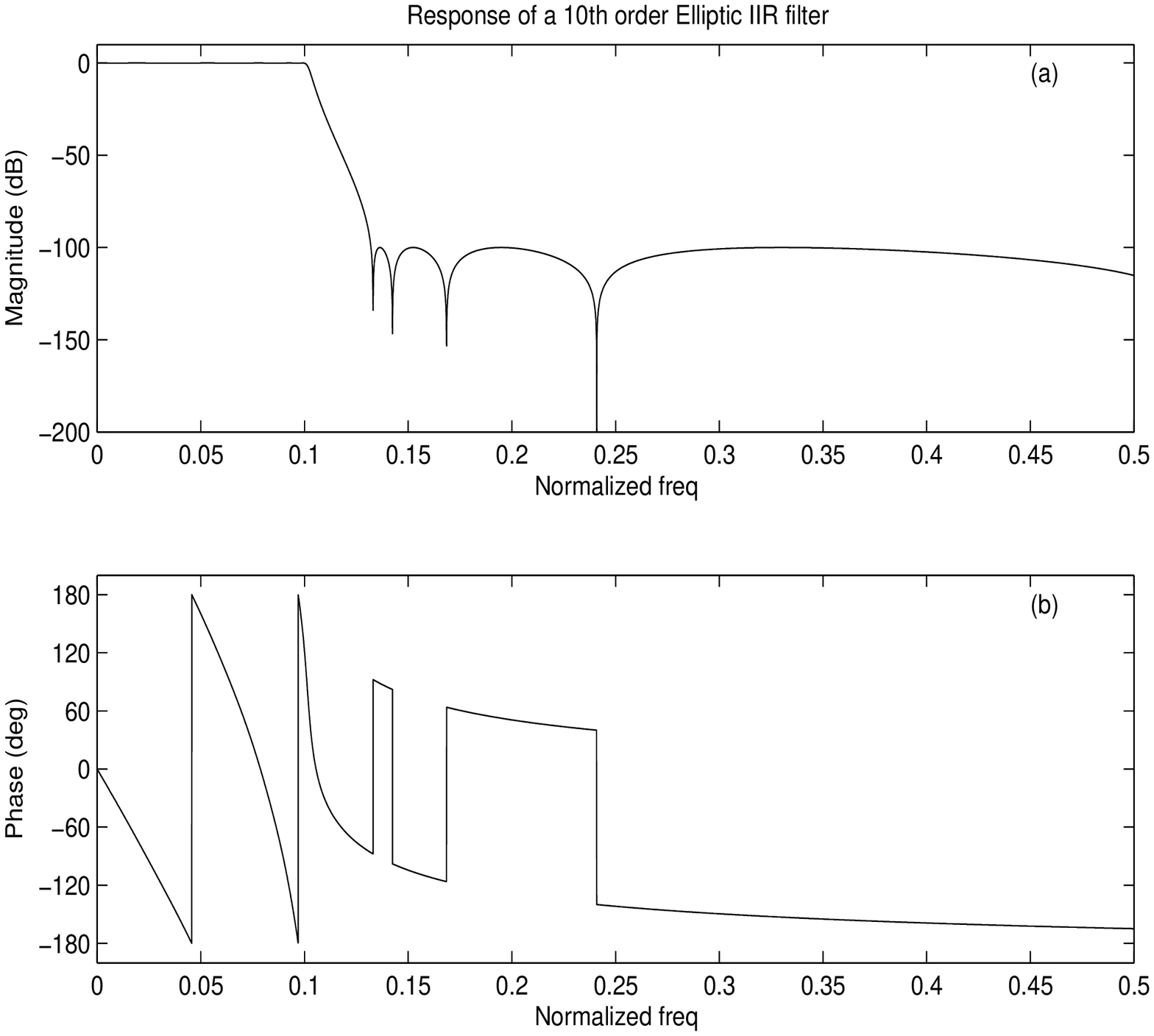}
\caption{Block diagram of a multirate filter (left). The first block reduces the
sampling frequency. The basic filter that needs to be implemented
follows the resampling block, which is a digital elliptic filter in our design.
The amplitude and phase response of the elliptic filter is shown on right.
\label{fig3}
}
\end{figure}

Present day correlators are designed using digital techniques.
The typical bandwidth of the interferometer is a few tens of MHz or more. 
Sampling the signals at Nyquist rate implies that the data rate at the output of the multiplier
will be of the order of the bandwidth. As mentioned above the output of the
correlator can be averaged for several secs. This means that the required cutoff frequency
of the filter that replaces the simple averaging should be a few tenths of a Hz. 
Thus the normalized (normalized with half the Nyquist frequency of the baseband) cutoff frequency  of the
filter is $< 10^{-6}$. A direct realization of a digital filter is not practical in such 
situations. However, such filters are realizable
using multirate filtering technique. This technique essentially implements
the required filter after reducing the sampling frequency
(see Fig. 3; Bellanger 2000). 

The basic filter used for the multirate filtering design is
a digital elliptic filter. Using Matlab, a digital elliptic filter 
of order 10 with normalized cutoff frequency of 0.1
and stopband attenuation of 100 dB is designed. The passband
ripple is chosen as 0.1 dB. The elliptic filter is realized using IIR 
(infinite impulse response) filter sections and its response is
shown in Fig. 3. The sampling rate 
reduction is done in two ways: (1) using polyphase
filtering technique; (2) by low-pass filtering the data using a moving
average filter and then resampling the data at half the number of points
averaged. The second type of filter is economical and simpler to
implement in hardware. For testing the performance of the filter the
sampling rate is reduced by 100. Thus the effective cutoff 
frequency of the filter is $10^{-3}$ and we believe that this could
be scaled further down by appropriately resampling the data.
The responses of the multirate filters are shown in Fig. 4. 

\begin{figure}
\plotone{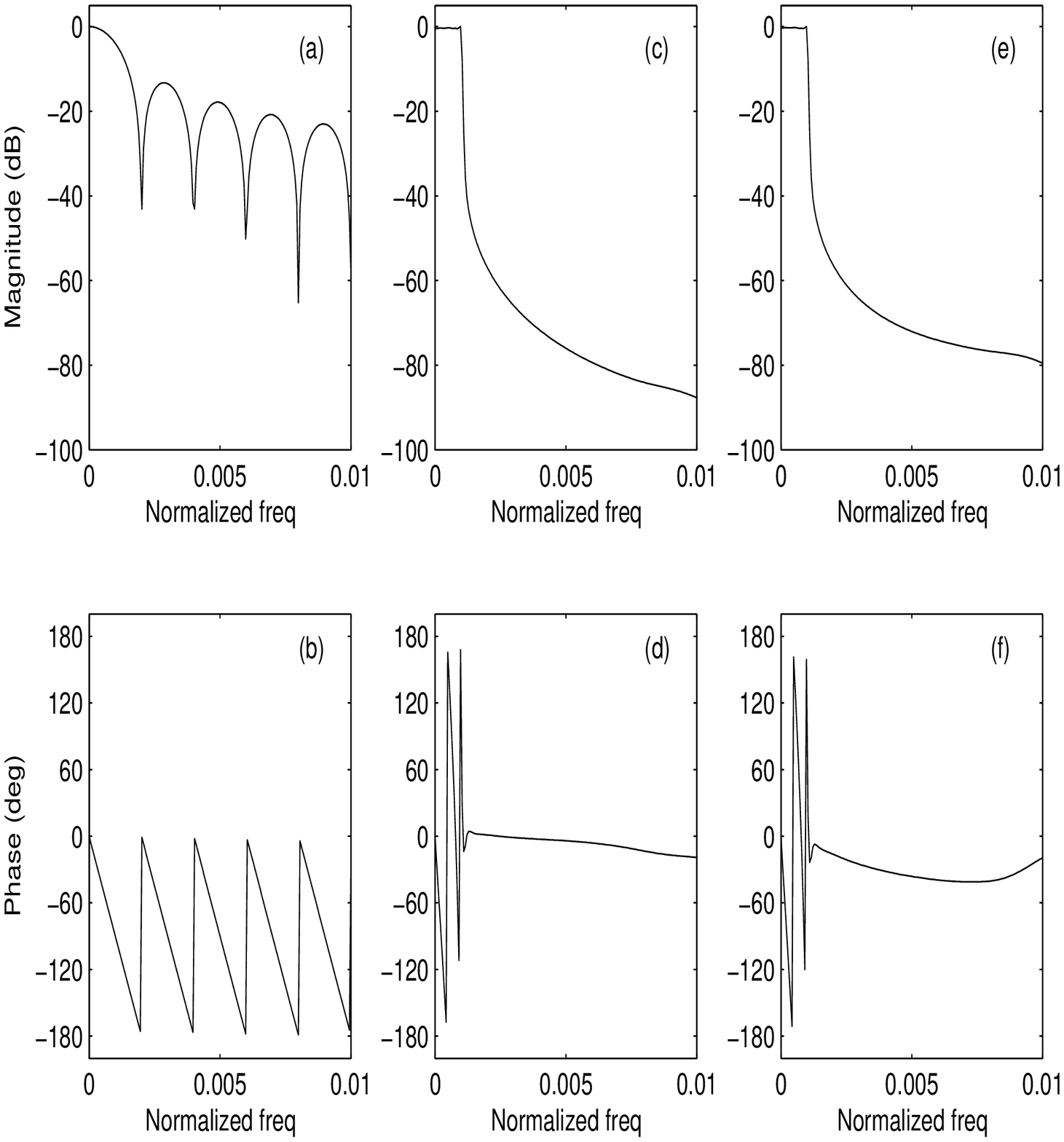}
\caption{Multirate filter responses along with that of a moving average
filter. The amplitude and phase response of a 1000 point moving average
filter is shown in (a) \& (b). Response of a multirate filter with
resampling done using polyphase filtering  followed by a digital elliptic
filter is shown in figure (c) \& (d). Figs. (e) \& (f) shows the
response of a multirate filter with a moving average filter followed
by an elliptic filter. (See text for details). 
\label{fig4}
}
\end{figure}
   
\subsection{Results of the simulation}

\begin{figure}
\plotone{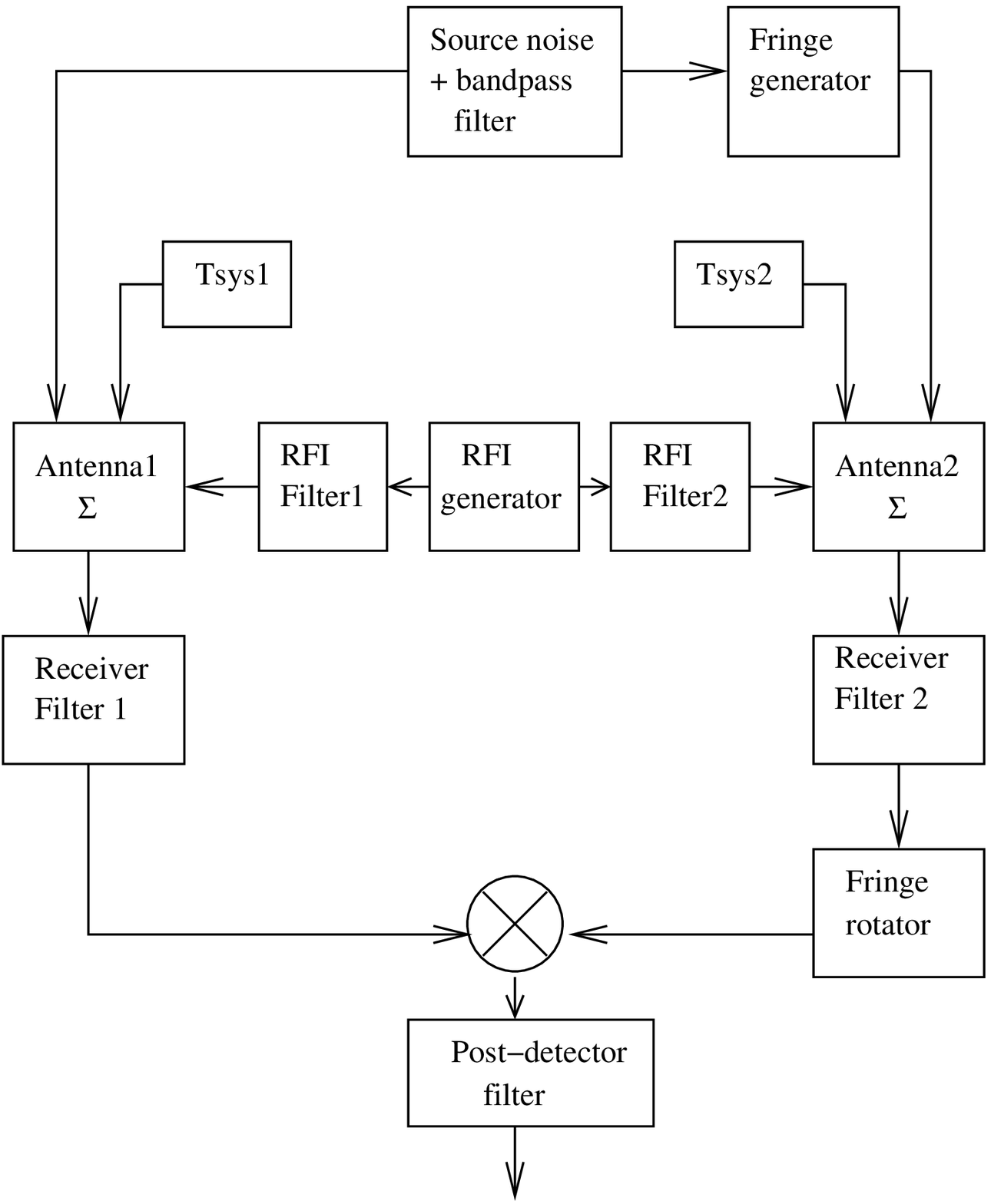}
\caption{Block diagram of the two element interferometer used
for the simulation. The fringe generator introduces a fringe
in the source noise and the fringe rotator block does the `fringe stopping'.
A moving average filter and two types of multirate filters are used
as post-detection filters (see Sec. 3) to determine their relative 
performance in suppressing RFI.
\label{fig5}
}
\end{figure}

A simulation was done in Matlab to determine the improvement in RFI rejection
of a two element interferometer by implementing the multirate filtering process. 
Fig. 5 shows the block diagram of the interferometer 
implemented for the simulation. The fringe generator
adds a fringe to the source noise. The antenna blocks are power combiners,
combining the source noise, the two independent system noises and the RFI.
All noises are Gaussian random variables. The response of the receiver is
realized using the receiver filters. Fringe stopping is done before
multiplying the two antenna outputs. The relative delay between the two 
antennas is made Zero for the simulation. 

\begin{figure}
\plottwo{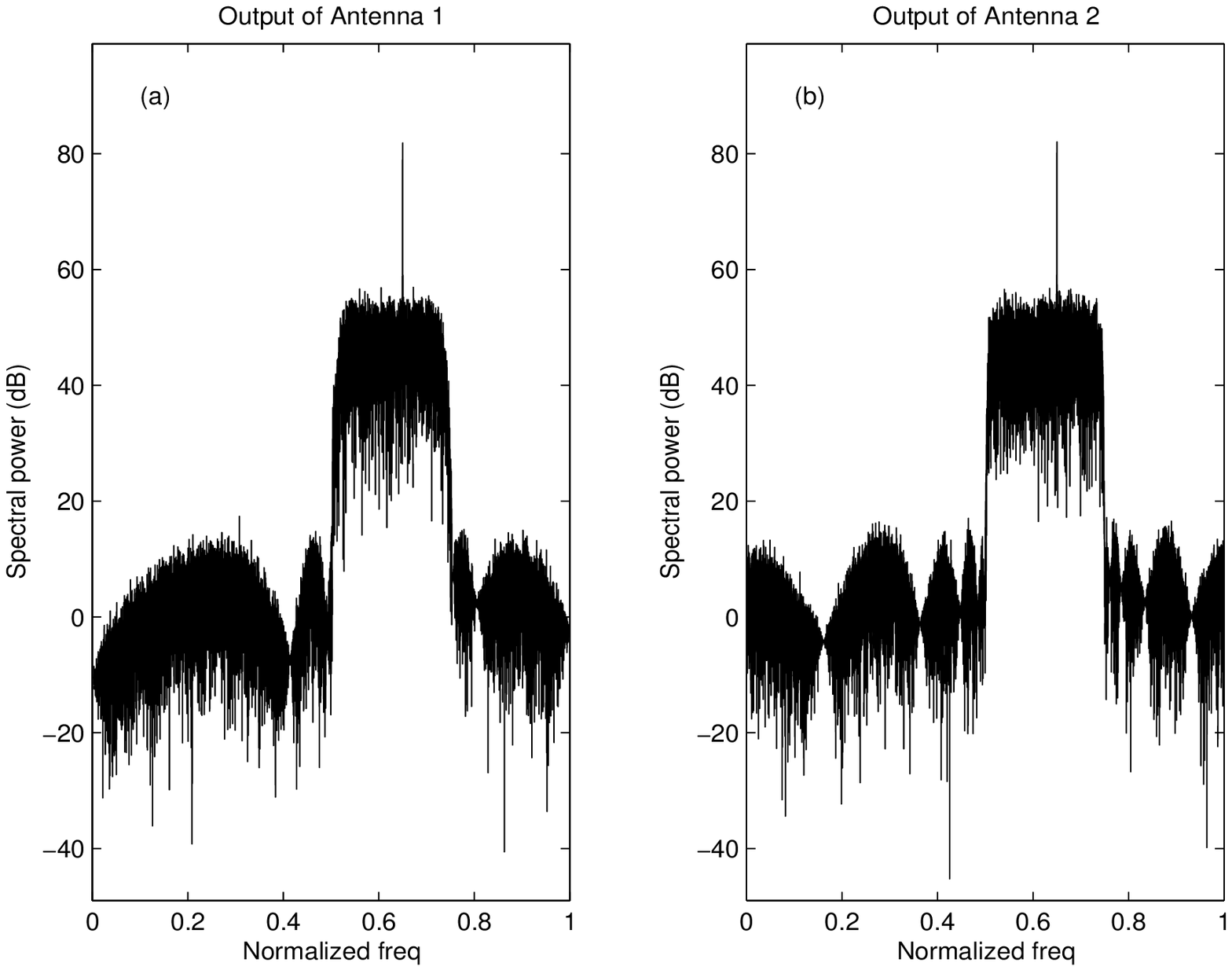}{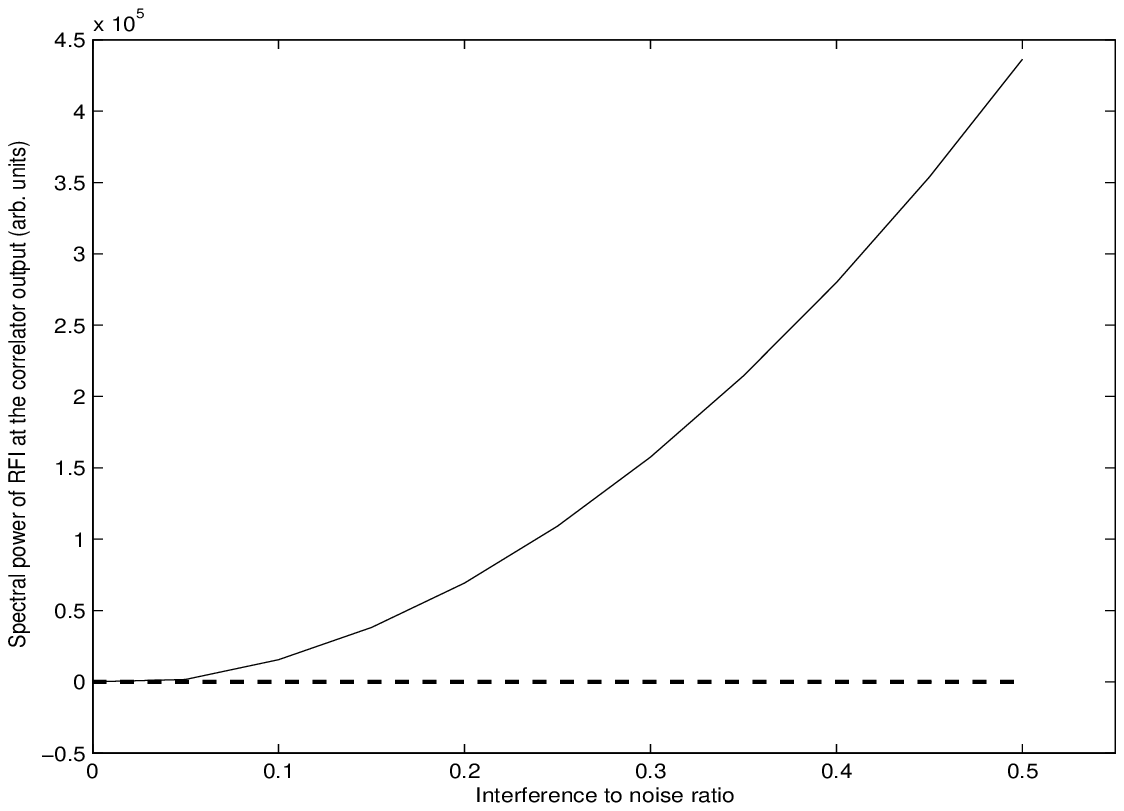}
\caption{Spectra of the outputs of the antennas obtained in the simulation is shown on left.
The RFI is located at 0.65. The spectral power of the RFI at the output of the correlator
against the interference-to-noise ratio at the antenna output is plotted on right. The data
with a moving average post-detection filter is shown in solid curve. The dashed line 
corresponds to the data with multirate filter using polyphase technique as post-detection filter. 
\label{fig6}
}
\end{figure}

\begin{figure}
\plotone{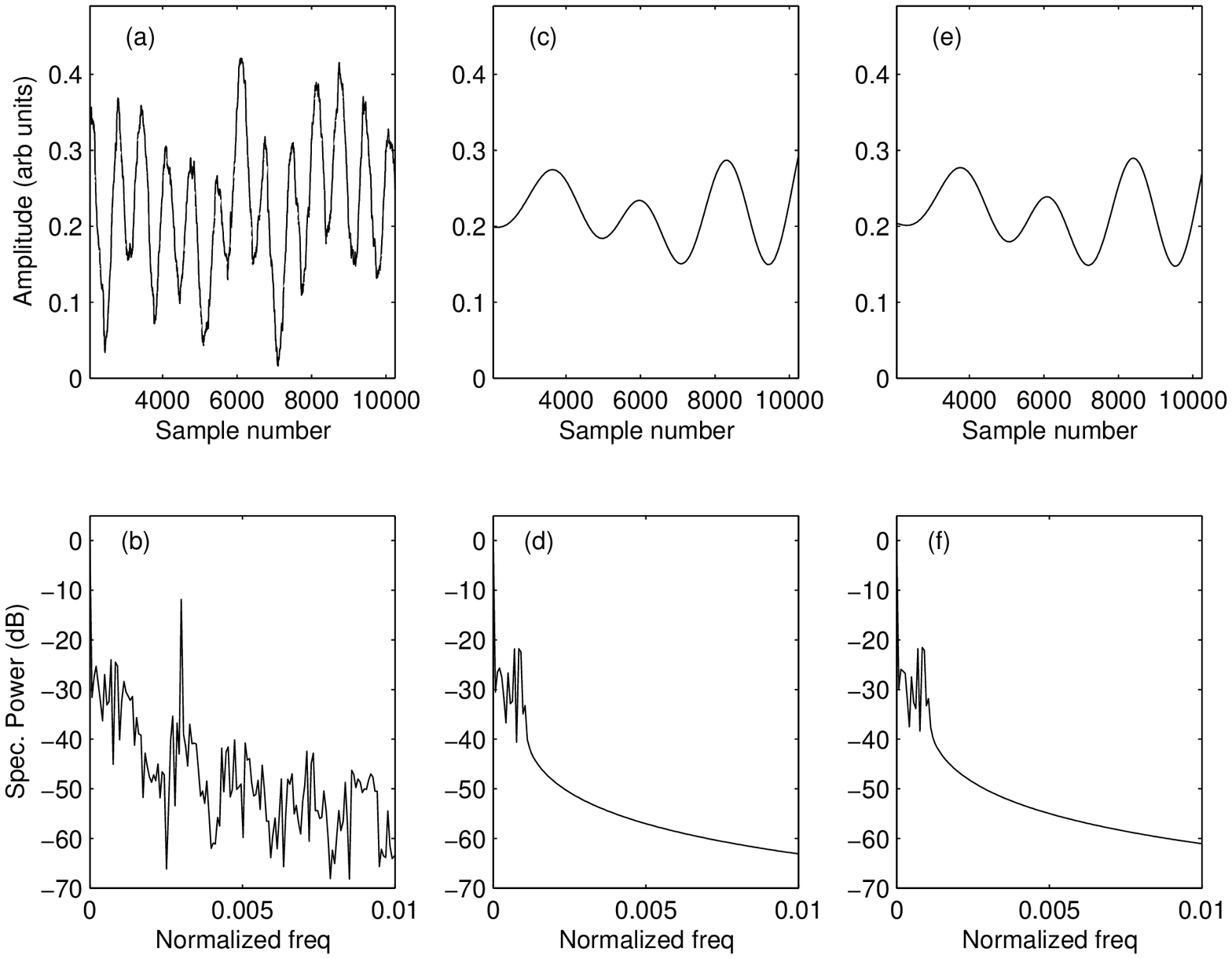}
\caption{Time series of the correlator output and its spectrum.
Figs (a) \& (b) are obtained with a moving average post-detection filter.
A multirate filter which uses polyphase filter to resample the data is used
as the post-detection filter to get Figs (c) \& (d). Figs (e) \& (f) are
obtained using a multirate filter which uses a moving average filter
to resample the data. 
\label{fig7}
}
\end{figure}

The designed multirate filters and the moving average filter 
are used as post-detection filters to study their relative performance.
A sine wave with normalized frequency of 0.65 is used 
as RFI. The receiver bandwidth in normalized frequency is selected as 0.2 
centered at 0.625. The receiver noise ($T_{sys}$) and source noise powers are
made equal to 0.2 in arbitrary units. Fig. 6 shows the spectra of the
two receiver outputs. The source noise is multiplied with a fringe of normalized 
frequency $3 \times 10^{-3}$. If we consider that this fringe freq should
translate to say 3 Hz, then the Nyquist frequency will be 2.0 KHz. This choice
of the fringe frequency is limited by the practicality of the simulation.  
The outputs of the correlator for different types of post-detection
filters are shown in Fig. 7. The remarkable ($>$ 40 dB) improvement
in the RFI suppression compared to the case when a moving average filter
is used as post-detection filter is evident from this figure. 
Comparing the spectrum of the correlator output
with a moving average post-detection filter and multirate post-detection filter 
shows that the spectral
values below the cutoff freq of the filter are almost the same for both cases. 
This means that the amplitudes of the desired fringe frequencies are
not affected by the multirate filtering process.

The effectiveness of the technique is also determined by estimating the spectral power
of the RFI at the output of the correlator with different types of post-correlation filters. 
To estimate the spectral power of RFI a power spectrum of the
correlator output is first made. The difference of the spectral values
of the channels where the RFI is expected and the average of the values of the 
two nearby channels is then taken. If the RFI power is comparable to the noise
fluctuations in the spectrum then the difference can also be negative. We assigned
a value of zero for the RFI spectral power if the estimate is negative. 
Fig. 6 shows the spectral power of RFI vs the interference-to-noise
ratio (INR) of the RFI at the output of the receiver. The INR
is the ratio of the RFI power to the noise power (receiver noise + source noise) at 
the output of the receiver. The remarkable suppression of the spectral power of 
the RFI with a multirate filter in post-detection 
stage is clear from Fig. 6.   

\section{Limitations of the technique}
\label{seclim}

The technique discussed in this paper to improve RFI suppression of interferometers
is effective only for correlated RFI (see Sec. 2). Usually terrestrial RFI
is correlated on baselines of length less than a few 100 kms. Correlated RFI produces
fringes at the output of the interferometer. The fringe frequency depends on the 
baseline length as well as the source declination for a given hour angle.
Effective suppression of RFI occurs if the fringe frequency is much above
the cut-off frequency of the multirate post-detection filter (see Sec. 3). 
However, if the baselines are small ($<$ a few kms) then the fringe frequency will 
be small and it becomes difficult to make filters that can attenuate these
frequencies.  Thus the technique presented here gives good RFI rejection for 
intermediate ($\sim$ a few kms to a few 100 kms) baselines. The fringe frequency also 
becomes smaller for sources near the north celestial pole (declinations near 90\deg),
which means that this technique will not work efficiently for this case as well.

Often RFI will have a strong carrier and associated modulating frequency components. 
The modulation results in a wider spectrum of the RFI at the output of the 
correlator. Thus only those components below the cut-off frequency of the 
multirate filter will be attenuated.

The analysis and simulation presented in this paper consider that the RFI is not time
variable. However, in reality RFI is time variable. The response of the multirate filter to
transient RFI needs to be studied. This will be presented elsewhere.

\section{Effects of multirate filtering on the visibility and spatial domain}
\label{secvis}

Till now the effect of the multirate filtering technique is discussed in relation
with the time series at the output of the correlator. Any processing of 
the correlator output has an equivalent effect on the visibility domain 
and correspondingly on the spatial domain (i.e. positions in sky). The 
effective position of a stationary RFI is the north celestial pole. Thus
suppressing the RFI using the multirate filtering process is 
equivalent to introducing a spatial null at the north celestial pole.
In fact, the spatial null is not confined to the north celestial pole alone
but extends along the spatial domain where the differential fringe
frequencies are larger than the cutoff frequency of the multirate filter. 
The attenuation due to the spatial null will be
proportional to the attenuation of the RFI due to the multirate filter.
 
In the visibility domain, the multirate filtering process essentially convolves the
true visibility with the filter response along $uv$ tracks traced  
by pairs of antennas in an interferometer. Depending on the filter response, the
amplitude and phase of the spatial frequencies of interest are modified.  
Since the filter is implemented digitally its response
is known well and is identical for all the correlators. With
the knowledge of this filter response, we believe a correction similar to the
primary beam correction can be done to take out the effect of filtering. 
A detailed discussion on this will be presented elsewhere.  

\section{Acknowledgment}
DAR thanks Rick Fisher for all useful discussions during the course of this work.
The basic idea described in this paper stemmed from a discussion DAR had with Prof. A. Pramesh
Rao, National Center For Radio Astrophysics, Pune, India.

\begin{question}{B.\ Carlson}
Could notch filters be used instead?
\end{question}
\begin{answer}{Anish Roshi}
No, using notch filter instead of low-pass filter to suppress the fringe
due to RFI has two disadvantages. (a) Since the fringe frequency is a 
function of baseline length the characteristics of the required notch 
filter for each baseline is different. This difference in filter 
characteristics can introduce baseline based effects thus may increase
closure errors. (b) Man made RFI in general are modulated carriers. The
modulation results in amplitude modulation of the fringe due to the 
interference. Using a low-pass filter, instead of notch filter, 
suppresses most of the sidebands of the `modulated fringe' thus 
suppressing the RFI better.
\end{answer}

\begin{question}{D.\ Ferris}
Replacing the normal rectangular integration function with the impulse
response of your filter can be expected to increase the variance of the
data, as well as smearing along time tracks in the u-v plane.
Do you have estimates of these effects for the filters you presented?
\end{question}

\begin{answer}{Anish Roshi}
Regarding variance, there will not be an increase in variance
due to multirate filtering if the noise equivalent
bandwidth of the filter is smaller than that of the conventional 
averaging filter. Regarding the smearing in the u-v plane: in the
conventional analysis (see Thompson et. al 2001, p205) of the 
smearing of visibility in the u-v plane due to averaging,
the averaging time is determined such that the u-v track in the
u-v plane should not cross the cell size in the u-v plane, which
is determined by the maximum field-of-view to be mapped. The 
condition thus obtained
on the integration time is identical to that given by Eq. (5) in the
text. The integration time is just the inverse of the maximum differential
fringe frequency in Eq (5). Thus, as long
as the multirate filter retains the maximum differential fringe frequency 
given by Eq. (5), there will not be any smearing in u-v
plane for those sources within the field-of-view.
\end{answer}

\end{document}